\documentstyle[aps,twocolumn,prl,psfig]{revtex}

\begin{document}
\draft
\title{Non-linear electrical conduction and broad band noise in 
the charge-ordered rare earth manganate Nd$_{0.5}$Ca$_{0.5}$MnO$_3$ }
\author{Ayan Guha\footnote[1]{email: aguha@physics.iisc.ernet.in}, Arindam Ghosh, A.K.Raychaudhuri }
\address{Department of Physics, Indian Institute of Science, 
Bangalore 560 012, India}
\author{S. Parashar, A.R.Raju, C.N.R.Rao}
\address{Jawaharlal Nehru Centre for Advanced Scientific Research, 
\mbox{Jakkur P.O., Bangalore 560 064, India}}

\twocolumn[\hsize\textwidth\columnwidth\hsize\csname 
@twocolumnfalse\endcsname

\maketitle
\begin{abstract}
Measurements of the dc transport properties and the  low-frequency
conductivity noise in  films of charge ordered Nd$_{0.5}$Ca$_{0.5}$MnO$_3$ 
grown on Si subtrate reveal the existence
of a threshold field in the charge ordered regime beyond which strong
non linear conduction sets in along with a large broad band conductivity noise.
 Threshold-dependent conduction disappears
as $T \rightarrow T_{CO}$,the charge ordering temperature.
This observation suggests that the charge ordered state gets depinned at the onset of 
the non-linear conduction.

\end{abstract}
\pacs{}

]

\newpage
Rare-earth manganites with 
a general chemical formula Re$_{1-x}$Ae$_x$MnO$_3$( where Re is a 
trivalent rare-earth and Ae is a divalent alkaline earth cation) show
a number of interesting phenomena like Colossal Magnetoresistance (CMR) 
and Charge Ordering (CO)\cite{RAO}.  
 These compounds belong to the ABO$_3$ type perovskite oxides
where Re and Ae ions occupy the A site and Mn occupies the B site. It has 
been known for some time that these manganites (depending on the size of
the average A-site cationic radius   \mbox{$< r_A >$} ) can charge 
order, for certain values of x.
The nature of the CO state depends on the value of \mbox{$< r_A >$} 
and it is stabilized if the value of \mbox{$< r_A >$} is smaller.The CO transition is 
associated with a lattice distortion as well as orbital and spin ordering.

\vspace {0.5cm}

Recent experiments have established that CO state is strongly destabilized
 by a number of different types of
perturbations. An applied magnetic field of sufficient magnitude can lead
to a collapse of the CO gap $\Delta_{CO}$ and melting  of the CO 
state~\cite{TOKURA1,AMLAN}. The CO phenomenon is stabilized by lattice distortion. A 
perturbation to the distortion can also destabilize the CO state~\cite{AKR1}. Recently
it has been reported that application of an electric field~\cite{TOKURA2,SACHIN}, optical radiation~\cite{TOKURA3},
or x-ray radiation~\cite{TOKURA4} melts the CO state in Pr$_{0.7}$Ca$_{0.3}$MnO$_3$. It is
not clear, however, what causes destabilization of the CO state in these
cases and whether the underlying mechanism is same for all perturbations.

\vspace {0.5cm}

Electric field induced melting of the CO state leads to a  strong non-linear 
conduction as seen in the bulk~\cite{TOKURA2} as well as in films~\cite{SACHIN}. This raises a very 
important question whether there is a threshold field associated with the 
non-linear conduction.
In a driven system pinned by a periodic potential there exists a threshold 
force beyond which the system is depinned~\cite{GRUNER}. If the system is charged and the
driving force comes from an electric field then this shows as a threshold
field or bias for the onset of a non-linear conduction.Existence of a 
threshold field would imply that the melting of the CO state by an applied 
electric field can actually be a depining phenomena.
We investigated this in  films
of the CO system Nd$_{0.5}$Ca$_{0.5}$MnO$_3$ by careful measurement of field
dependent dc transport at various temperatures and also followed it up with a
 measurement of electrical
noise (voltage fluctuation) as a function of applied dc bias. We made the 
following important observations : \\
(1) There indeed exists a threshold field ($E_{th}$) below the CO          
temperature $T_{CO}$ and 
for $E > E_{th}$ a strong nonlinear conduction sets in.\\
(2) $E_{th}$ strongly depends on $T$ and $E_{th} \rightarrow$ 0 as 
$T  \rightarrow T_{CO}$.\\
(3) For \mbox{$T < T_{CO}$}, a large voltage fluctuation \mbox{($<\delta V^2 > / V^2$)} 
appears at the threshold field. Both $E_{th}$ and \mbox{$<\delta V^2 > / V^2$} 
reaches
a maximum at $T \approx$ 90K ($\approx 0.4T_{CO}$).\\
(4) The spectral power distribution of the voltage fluctuation is broad
band and has nearly 1/f character.

\vspace {0.3 cm}

In Nd$_{0.5}$Ca$_{0.5}$MnO$_3$, a system with relatively small $<r_A>$, the
CO transition takes place from a high temperature charge disordered insulating 
phase to a charge ordered insulating phase (COI). Charge ordering in this
system has been studied by us in details previously~\cite{AKR2}.
Poly-crystalline films of Nd$_{0.5}$Ca$_{0.5}$MnO$_3$ (average 
thickness$\approx$ 1000 nm) were deposited on Si(100) single 
crystal substrates by nebulized spray pyrolysis of organometallic compounds.
 The details of sample preparation and characterisation (including X-ray) 
have been given elsewhere~\cite{SACHIN}.
Contacts were made by sputtering gold on the films and 
 connecting the current and voltage leads on the gold contacts by silver   
paint. The I-V characteristics was measured by  dc current biasing and the 
 voltage between the voltage leads was 
measured by a nano-voltmeter . For measuring 
the electrical noise,  the fluctuating component
of the voltage $\delta V$ was amplified by
 $5 X 10^3$ times by a low noise pre-amplifier. The output of the 
pre-amplifier was sampled  by an ADC card and the data 
were directly transferred to the computer. The temperature was controlled
to within 10 mK for both the measurements.\\

\vspace {0.3 cm}

The films have a $T_{CO}$$\approx$250K as seen from the resistivity        
data.The resistivity was measured at a measuring current of 3 nA, which is 
much lower than the current where non-linear conductivity sets in. The 
experiment was conducted down to 80K where the sample resistance becomes 
more than 100M$\Omega$, the limit of our detection electronics.
 
\vspace {0.3 cm}

In figure 1, we show the typical I-V curves at few characteristic 
temperatures.At all the temperatures (except that at 220K) there is a 
clear signature of a threshold voltage $V_{th}$ beyond which the current
rises significantly signalling the onset of strong 
nonlinear conduction. (The separation of electrodes in our experiment is
2 mm, so that $E_{th} = 5 V_{th}$ volts/cm). I-V curves show 
two components of conduction: a normal component which exists at all V and a 
strongly non-linear component starting at  $V > V_{th}$. 
 The normal component although not exactly linear in I-V, has much less
non linearity. We fit our I-V data using the following empirical expression
which allows us to separate out the two components :\\
\begin {equation}
I = f_{1}(V - V_{th}) + f_{2}(V) \\
  = C_{1}(V - V_{th})^{n_{1}} + C_{2}V^{n_{2}} \\
\end {equation}
\noindent where $f_1$, a function of $(V - V_{th})$, is the component of
current that has a threshold associated with it and $C_{1} = 0$ for $V < 
V_{th}$.
The component $f_{2}$ is the normal conduction component.
 $C_{1}$,$C_{2}$,$n_{1}$,$n_{2}$ are constants for a given
temperature. The data at all temperatures can be well fitted to eqn.1 for 
$T >$ 90K as shown by the solid lines in figure 1. The dashed and 
dashed-dotted lines give the contributions of each of the terms. For $T <$ 90K
certain additional features show up (see data at 81K) in the I-V data 
which give impression that
there may be multiple thresholds. In figure 2(a) we have plotted 
the threshold voltage $V_{th}$ as a function of $T$ as obtained from eqn.1. 
It can be seen that $V_{th} \rightarrow 0$ as $ T \rightarrow T_{CO}$. 
Within the limitations of our detectibility, we could see a finite 
nonzero $V_{th}$ upto $T\approx 170K \approx 0.7 T_{CO}$.Beyond this       
temperature it is difficult to distinguish between the two conduction       
components.

\vspace{0.5cm}

The relative contributions of $f_{1}$ and $f_{2}$
 to the total current (expressed as the ratio $f_{1} / f_{2}$ 
evaluated at $I = 1 \mu A$) has been plotted as a function of T in     
figure~2(b). At T<<T$_{CO}$, the non-linear component is orders of 
magnitude larger than the normal conduction component and they are 
comparable as T$\rightarrow T_{CO}$.
 The exponent $n_1$ is strongly 
temperature dependent and from a value $\approx 2$ at 160K it reaches a value more
than 5 at $T \approx 100K$. The exponent
$n_2$ does not have much of a temperature dependence and is $\approx 1.1- 1.4$
for T$\leq180K$.

\vspace{0.5cm}

In pinned driven system one often sees onset of broad band noise as the system 
is depinned at the threshold voltage~\cite{GRUNER}. We find that such is indeed the case
in this system. In figure 3 we show the magnitude of the voltage 
fluctuation$<\Delta V^2> / V^2$ as a function of the applied bias V at $T = 100K$ along 
with the I-V curve. The arrow indicates $V_{th}$. It is clear that the voltage
fluctuation has a non monotonous dependence on V and reaches a peak at 
$V \approx V_{th}$. This fluctuation has been seen at all $T < 0.7 T_{CO}$
where we can detect measurable $V_{th}$. The peak values of the 
fluctuation measured at different T are shown in figure 2(c). The fluctuation 
$ \rightarrow 0$ as $T \rightarrow T_{CO}$ and has a peak at 90K where     
$V_{th}$ also shows a peak.

\vspace{0.5cm}

Frequency dependences of the spectral power $S_{V}(f)$ 
measured at 100K with biases V<V$_{th}$, V$\approx$V$_{th}$ and V>V$_{th}$
are shown in figure 4 .
 We have plotted the data as $f.S_{V} / 
V^2$ vs. $f$. For a pure 1/f noise ($S_{V}$ $\propto$ 1/f), this should be 
a straight
line parallel to the f-axis. It can be seen that the predominant contribution to noise has
 1/f character.In addition,there is 
 another broad band contribution riding on the main 1/f
contribution. At
higher V the spectra becomes more of 1/f nature.

\vspace{0.5 cm}  
The onset of strong non-linear conduction at a threshold voltage and the 
accompanied broad band noise has been seen in solids like 
NbSe$_{3}$,TaS$_{3}$ which show depinning of charge density waves (CDW) 
by a threshold field~\cite{GRUNER}. Though the physics of CDW and CO states are entirely 
different, the underlying phenomenological description of depining can be 
similar. Electron diffraction (ED) and electron microscopy studies on a CO system 
(La$_{0.5}$Ca$_{0.5}$MnO$_3$) have shown that the CO is associated
with formation of stable pairs of Mn$^{3+}$O$_6$ stripes. The Mn$^{3+}$O$_6$
octahedra in the stripes are strongly distorted by the Jahn-Teller (JT) 
distortion~\cite{CHEONG}.It is possible that
 the stability of the CO system depends on the stability of the 
stripes which can be pinned. The strong JT distorted pairs of the 
Mn$^{3+}$O$_6$ octahedra can  act as periodic pinning sites due to 
local strain field. From our data for $T <$ 90K, it seems there are changes 
occuring below 90K. We are not clear about the changes. We only note that 
in magnetic studies we found that strong irreversibility sets in below 80K~\cite{AKR2}.

\vspace{0.5 cm}
To conclude, the present study demonstrates that there is a threshold field           
associated with the onset of non-linear conduction in the CO system 
along with the existence of a  broad    
band noise. The observation is taken as 
evidence of depinning of CO state as the origin of non-linear conduction in 
these solids.

\newpage
{\centerline {FIGURE CAPTIONS}}            

(1) FIG.~1. I-V curves at different temperatures, solid line shows the total I, 
dashed and dashed-dotted lines show the components f$_1$ and f$_2$.

(2) FIG.~2 Temperature variation of (a) resistivity, (b) magnitude of threshold
voltage, (c) relative contributions of f$_1$ and f$_2$ and (d) noise magnitude
at the threshold voltage

(3) FIG.~3 The noise magnitude and I-V characteristics at 100K. The arrow 
indicates the threshold voltage.

(4) FIG.~4 Frequency spectrum of the noise at 100K for different bias values.

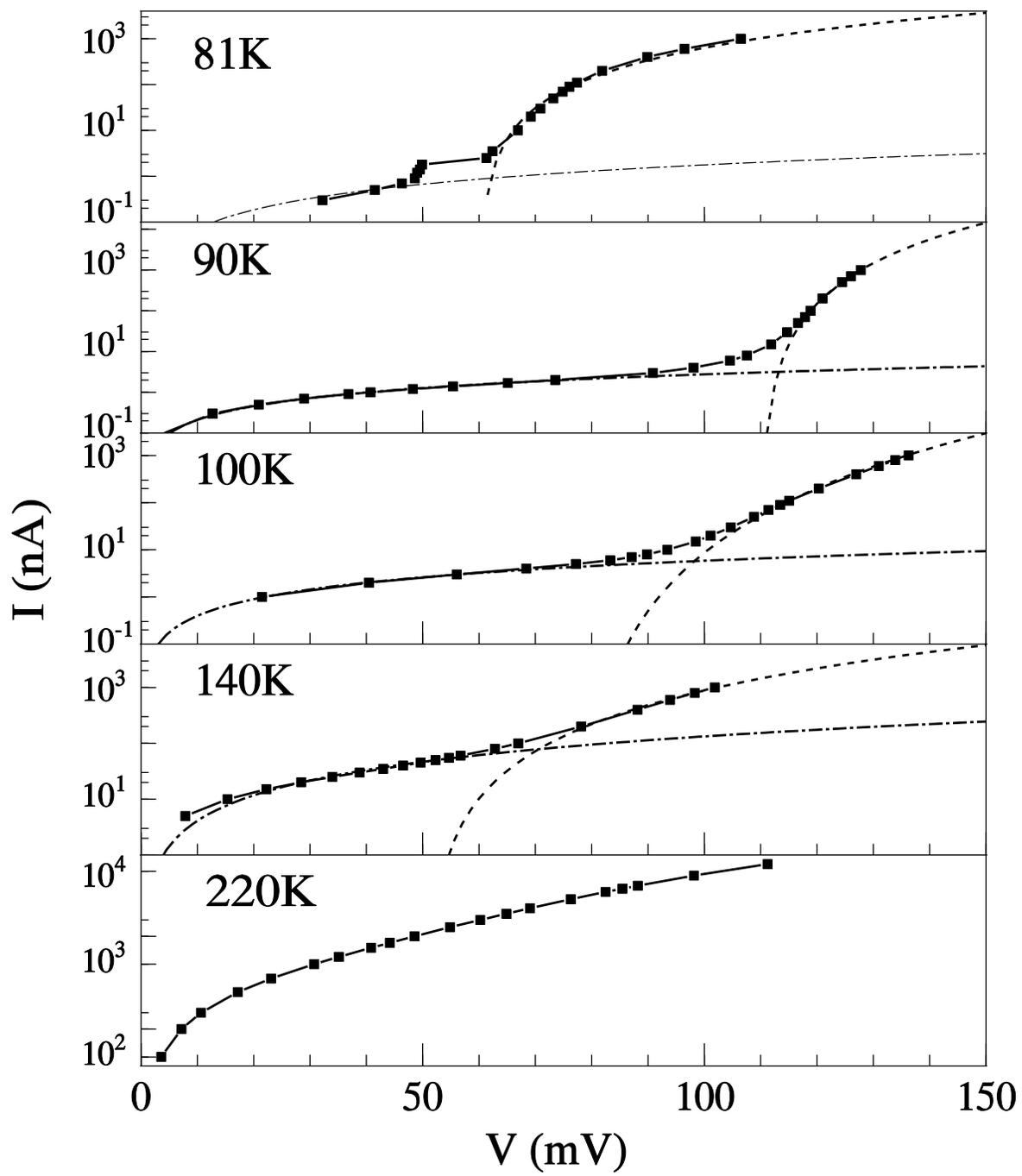
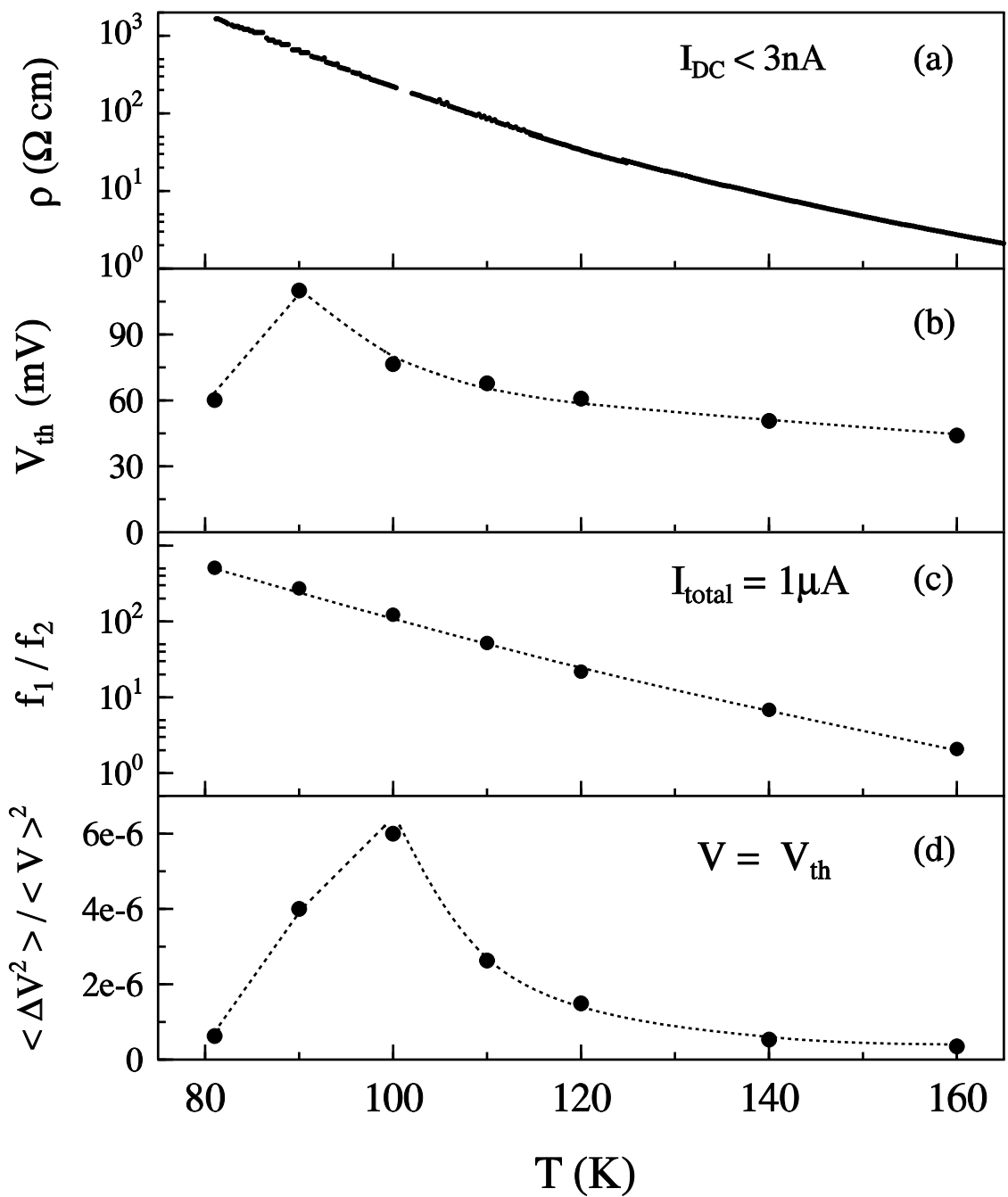
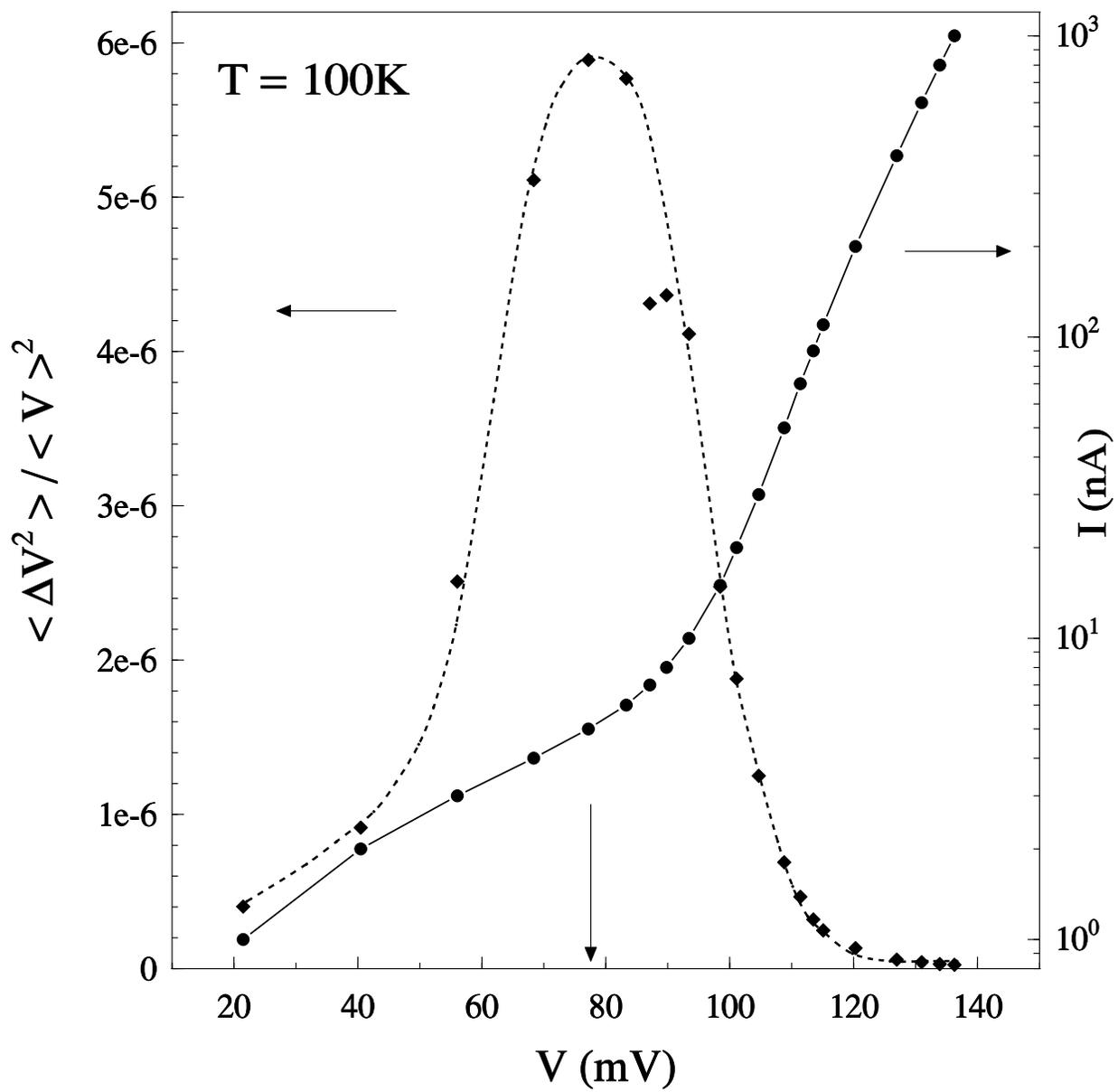
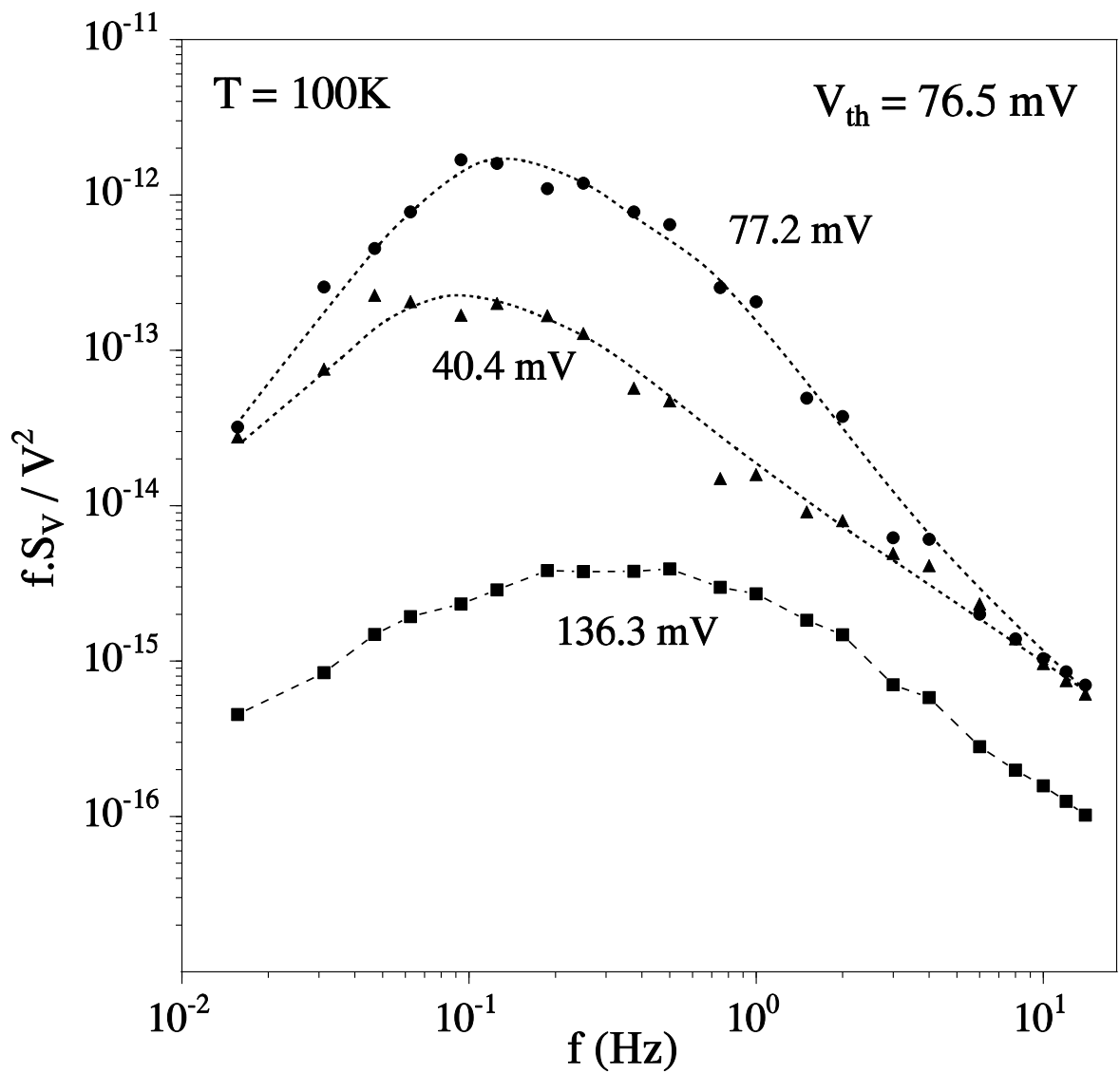
\end{document}